\documentclass[aps,prl,reprint]{revtex4-2}
\pdfoutput=1
\usepackage{graphicx}  
\usepackage{dcolumn}   
\usepackage{bm}        
\usepackage{amssymb}   
\usepackage{color,epsfig}
\usepackage{amsmath}
\usepackage{bigints}

\def\red#1{\textcolor{red}{#1}}
\def\blue#1{\textcolor{blue}{#1}}

\begin{document}
\newcommand{\HH}{H$_2$}
\newcommand{\pHH}{\emph{para}-H$_2$}
\newcommand{\oHH}{\emph{ortho}-H$_2$}
\newcommand{\Eref}[1]{Eq.~(\ref{#1})}
\newcommand{\Fref}[1]{Fig.~\ref{#1}}
\newcommand{\etal}{\emph{et al.}}

\def\red#1{\textcolor{red}{#1}}
\def\blue#1{\textcolor{blue}{#1}}


\title{Stereodynamical control of cold collisions between two aligned D$_2$ molecules}
\author{Pablo G. Jambrina} \affiliation{Departamento de Qu\'{\i}mica F\'{\i}sica. Universidad
de Salamanca, Salamanca 37008, Spain}\email{pjambrina@usal.es}
\author{James F. E. Croft} \affiliation{The Dood-Walls Centre for Photonic and Quantum
Technologies, Dunedin, New Zealand} \affiliation{Department of Physics, University of Otago,
Dunedin, New Zealand}\email{j.croft@otago.ac.nz}
\author{Junxiang Zuo}\affiliation{Department of Chemistry and Chemical Biology, University
of New Mexico, Albuquerque, New Mexico 87131, USA}\email{hguo@unm.edu}
\author{Hua Guo}\affiliation{Department of Chemistry and Chemical Biology, University of
New Mexico, Albuquerque, New Mexico 87131, USA}\email{hguo@unm.edu}
\author{Naduvalath Balakrishnan}\affiliation{Department of Chemistry and Biochemistry,
University of Nevada,
Las Vegas, Nevada 89154, USA}\email{naduvala@unlv.nevada.edu}
\author{F. Javier Aoiz} \affiliation{Departamento de Qu\'{\i}mica F\'{\i}sica. Universidad
Complutense.
Madrid 28040, Spain}\email{aoiz@quim.ucm.es}
\date{\today}

\begin{abstract}

Resonant scattering of optically state-prepared and aligned molecules in the cold regime
allows the most detailed interrogation and control of bimolecular collisions. This technique
has recently been applied to collisions of two aligned ortho-D$_2$ molecules prepared in
the $j=2$ rotational level of the $v=2$ vibrational manifold using the Stark-induced
adiabatic Raman passage technique. Here, we develop the theoretical formalism for
collisions of two aligned molecules and apply our approach to state-prepared
D$_2(v=2,j=2)$+ D$_2(v=2,j=2)\to$ D$_2(v=2,j=2)$+ D$_2(v=2,j=0)$ collisions. Quantum scattering calculations were performed  in
full-dimensionality on an accurate H$_2$-H$_2$ interaction potential. Key features
of the experimental angular distributions are reproduced and attributed primarily to a
partial wave resonance with orbital angular momentum $\ell=4$.

\end{abstract}

\pacs{}
\maketitle

\section{introduction}

In molecular encounters collision outcomes are influenced by factors such as the collision
energy ($E_{\rm coll}$)  and directional properties (orientation and alignment). While   measurements of the
energy (actually the kinetic temperature, $T$) dependence of the collision rates are rather
routine, experiments that measure the dependence of the outcome of a molecular collision  on the initial
alignments (stereodynamics) of the reactants are scarce (see for example Refs.
\citenum{Wang2011,Wang2012,Wang2016a,Wang2016b,Brouard2015,Chadwick2014,Brouard2013,VKKBOAGM:NC18,OGVAKNAGBM:NC17,SLLMJACC:NC18,2017_Science_Perreault,2018_NatChem_Perreault,sarp_hed2,HWBJA:NC19,CPWJAB:JPCL21,HBWGJAB:PCCP20,WHJAB:JPCA19,sarp_hed2,sarp_hed2_science,SARP_HD-He}).

Optical state-preparation using the Stark-induced adiabatic Raman passage (SARP) method
combined with co-expansion of the colliding species has become
a versatile tool to explore stereodynamics of atom-molecule and molecule-molecule
collisions~\cite{2017_Science_Perreault,Chandika_JPLC_2017,Chandika_JCP_2020,2018_NatChem_Perreault,SARP_HD-He,sarp_hed2,sarp_hed2_science}.
When applied to light molecules such as HD and D$_2$, relative collision energies
near $\sim$ 1 K can be achieved,  as demonstrated for
HD+H$_2$/D$_2$~\cite{2017_Science_Perreault,2018_NatChem_Perreault},
HD+He~\cite{SARP_HD-He} and D$_2$+He~\cite{sarp_hed2,sarp_hed2_science} mixtures. In
this regime, isolated resonances control the collision outcome, and their strength sometimes
depends on the relative alignment between the two
partners~\cite{2018_PRL_Croft,2019_JCP_Croft,2019_PRL_Jambrina,Morita_He-HD,hehd_morita_ultracold,morita_hcl-h2,HHF_Jambrina20,Jambrina_PCCP_2021,Jambrina_JPCL_HeD2_2022},
so the SARP method provides a powerful technique to study and control stereodynamics of
bimolecular collisions. However, most of these studies involve atom + molecule collisions, and
those that deal with bimolecular collisions could only control the direction of the internuclear
axis of one of the colliding
partners~\cite{2017_Science_Perreault,2018_NatChem_Perreault}.

Very recently, Zhou
\textit{et al.}\cite{Zhou:NC22} reported results of the inelastic collisions between two aligned
$ortho-$D$_2$($v$=2,$j$=2) molecules, showing how the angular distribution of the scattered
products depends sensitively on the direction of D$_2$ internuclear axis with regard to the
scattering frame defined by $\bm{k}$ and ${\bm k}'$, the reactant-approach and
product-recoil directions. Further,  while not directly observed, key features of
the angular distribution are attributed to a
resonance caused by the orbital angular momentum $\ell=2$ near 1\,K in the incoming
channel whose properties  are predicted to be strongly influenced by the initial alignment of
the two molecules.

Previous theoretical treatments of the stereodynamics of bimolecular collisions considered only the
polarization of one of the collision
partners~\cite{AMHKA:JPCA05,2018_PRL_Croft,2019_JCP_Croft,2019_PRL_Jambrina,Jambrina_PCCP_2021,Guo:JPCL22}.
Here, we present the theoretical formalism for the angular distribution of scattered
products when
both reactants are polarized. Using full-dimensional \textit{ab-initio} quantum scattering
calculations on an accurate potential energy surface (PES)~\cite{ZuoH4PES}, we reproduce the
experimental angular distributions reported by Zhou \textit{et al.}\cite{Zhou:NC22}.
Agreement with experiments is only obtained when collisions involving two polarized
molecules (both in $v=2$) as well as one polarized  (in $v=2$) and one unpolarized
molecule (in $v=0$, also present in the beam) are considered. Our results  reveal that there
is an $\ell$=4 partial wave resonance whose contribution to the experimental angular
distribution is dominant in the 1.5--3.5~K collision energy range.

\section{methods}

Let us consider collisions involving two molecules A and B, each of them in a
pure rotational state $j_{_{\rm A}}$ and $j_{_{\rm B}}$ and that we can control the spacial
distribution of the internuclear axis of one of them (for example, A). In that case, the
state-to-state differential cross section (DCS) can be calculated as \cite{AMHKA:JPCA05}
\begin{equation}\label{dcsalphabeta}
{\rm d}\sigma(\theta|\beta,\alpha) = \sum_{k=0}^{2j} \sum_{q=-k}^{k} (2 k + 1 )
\left[U^{(k)}_q(\theta)\right]^* a^{(k)}_q,
\end{equation}
where $a^{(k)}_q$ are the extrinsic polarization parameters that describe the anisotropic
preparation of the reactant in the $\bm k$--$\bm k'$ scattering frame.  If A is prepared in a
pure $|j_{_{\rm A}} m=0 \rangle$ state, where $m$ is the magnetic quantum number
determined with regard to a laboratory-fixed quantization axis (the polarization vector of the
Stokes and pump laser in the SARP experiment), the polarization parameters are given by
\begin{equation}\label{akq}
a^{(k)}_q = C_{kq} (\beta,\alpha) A^{(k)}_0 = C_{kq} (\beta,\alpha) \langle j_{_{\rm A}} 0, k 0 |
j_{_{\rm A}} 0  \rangle,
\end{equation}
where $A^{(k)}_0$ are the extrinsic polarization parameters in the laboratory frame, $
C_{kq}$ are the modified spherical harmonics, whose arguments $\beta$ and $\alpha$ are
the polar and azimuthal angles that define the direction of the polarization vector in the
scattering frame, and $\langle ..,..|.. \rangle$ is the Clebsch-Gordan coefficient.  For an
isotropic internuclear axis distribution, the only non-zero $a^{(k)}_q$ element is $a^{(0)}_0$.

In Eq. \ref{dcsalphabeta}, $U^{(k)}_q(\theta)$ are the intrinsic polarization dependent DCSs
(PDDCSs) of the \{${\bm k}$--${\bm j}_{_{\rm A}}$\!\!--${\bm k}'$\} three-vector correlations
that
describe how the collision outcome depend on the relative geometry of the reactants.
$U^{(k)}_q(\theta)$ can be expressed in terms of the scattering amplitudes  in the helicity
representation,
$f_{j'_{\rm A} m'_{\rm A} j'_{\rm B} m'_{\rm B} \,j_{_{\rm A}} m_{_{\rm A}} j_{_{\rm B}}
m_{_{\rm
B}}}(\theta) \equiv  F_{m'_{\rm A} m'_{\rm B} \,m_{_{\rm A}} m_{_{\rm B}}}(\theta)$, as:
\begin{eqnarray} \label{eq:jPDDCS}
\nonumber U^{(k)}_q(\theta) &=& \frac{1}{(2j_{_{\rm A}}+1) (2j_{_{\rm B}}+1)}
\sum_{\substack{m'_{\rm A}, m'_{\rm B} \\ m_{\rm  A}, m_{_{\rm B}}}} \,  F_{m'_{\rm A}
m'_{\rm B} \,m_{_{\rm A}} m_{_{\rm B}}}(\theta) \times
 \\ & & F^*_{m'_{\rm A} m'_{\rm B} \,(m_{_{\rm A}}+q) \, m_{_{\rm B}}}(\theta)  \langle
 j_{_{\rm A}} m_{_{\rm A}}, k q| j_{_{\rm A}} m_{_{\rm A}} + q\rangle \,,
\end{eqnarray}
with
\begin{eqnarray} \label{scatampl} \nonumber
F_{m'_{\rm A} m'_{\rm B} \,m_{_{\rm A}} m_{_{\rm B}}}(\theta) &=& \frac{1}{2 i k} \sum_{J}
(2J +1) d^J_{m'_{\rm A} + m'_{\rm B}, m_{_{\rm A}} + m_{_{\rm B}} }(\theta) \times \\ &&
S^J_{m'_{\rm A} m'_{\rm B} \,m_{_{\rm A}} m_{_{\rm B}}}(E),
\end{eqnarray}
where $d^J_{m'_{_{\rm A}} + m'_{_{\rm B}}, m_{_{\rm A}} + m_{_{\rm B}} }(\theta)$ is an
element of the Wigner reduced rotation matrix, and $S$ is an element of the Scattering
matrix in the helicity representation, with $m'_{_{\rm A}}$, $m'_{_{\rm B}}$, $m_{_{\rm
A}}$, and $m_{_{\rm B}}$ being the projections on $j'_{_{\rm A}}$, $j'_{_{\rm B}}$, $j_{_{\rm
A}}$, and $j_{_{\rm B}}$ on the initial and final relative velocities, respectively (the primed
indices are associated to the products states).

For two polarized reagents under the same polarization vector, the DCS can be expressed as
\begin{eqnarray}\label{dcsalphabeta4}\nonumber
{\rm d}\sigma(\theta|\beta, \alpha)  &=&  \sum_{k_{_{\rm A}}=0}^{2 j_{_{\rm A}}}
\sum_{q_{_{\rm A}}}
\sum_{k_{_{\rm B}}=0}^{2 j_{_{\rm B}}} \sum_{q_{_{\rm B}}} (2 k_{_{\rm A}} +1) (2 k_{_{\rm
B}} +1)  \\
&\times & \left[ U^{(k_{_{\rm A}},k_{_{\rm B}})}_{q_{_{\rm A}},q_{_{\rm B}}} (\theta)
\right]^*
a^{(k_{_{\rm A}})}_{q_{_{\rm A}}} a^{(k_{_{\rm B}})}_{q_{_{\rm B}}}
\end{eqnarray}
where each of the $a^{(k)}_q$ can be evaluated according to Eq.~\eqref{akq} as a function of
the
$\beta$ and $\alpha$ angles. The intrinsic \{${\bm k}$--${\bm j}_{_{\rm A}}$\!\!--$ {\bm
j}_{_{\rm
B}}$\!\!--${\bm k}'$\} 4-vector PDDCSs, $U^{(k_{_{\rm A}},k_{_{\rm B}})}_{q_{_{\rm
A}},q_{_{\rm
B}}}$, can be calculated as:
\begin{eqnarray}
&& U^{(k_{_{\rm A}},k_{_{\rm B}})}_{q_{_{\rm A}},q_{_{\rm B}}}(\theta) = \frac{1}{(2j_{_{\rm
A}}+1) (2j_{_{\rm B}}+1)} \times \\ \nonumber &&  \sum_{\substack{m'_{\rm A}, m'_{\rm B}
\\ m_{_{\rm A}}, m_{_{\rm B}}}} \,  F_{m'_{\rm A} m'_{\rm B} \,m_{_{\rm A}} m_{_{\rm
B}}}(\theta)
  F^*_{m'_{\rm A} m'_{\rm B} \,(m_{_{\rm A}}+q_{_{\rm A}}) (m_{_{\rm B}}+q_{_{\rm
  B}})}(\theta)  \times \\ \nonumber
  &&  \langle j_{_{\rm A}} m_{_{\rm A}}, k_{_{\rm A}} q_{_{\rm A}}| j_{_{\rm A}} (m_{_{\rm
  A}} + q_{_{\rm A}})\rangle  \langle j_{_{\rm B}} m_{_{\rm B}}, k_{_{\rm B}} q_{_{\rm B}}|
  j_{_{\rm B}} (m_{_{\rm B}} + q_{_{\rm B}})\rangle \,.
\end{eqnarray}
If either  $k_{_{\rm A}}$ or  $k_{_{\rm B}}$  is zero, we recover the three-vector PDDCS
$U^{(k)}_q(\theta)$. If $k_{_{\rm A}}=k_{_{\rm B}}=0$ we recover the $U^{(0)}_0(\theta)$,
the isotropic DCS.

The DCS in the SARP experiments that we aim to reproduce involves integration over the azimuthal angle ($\alpha$). This allows us to simplify the equation
\eqref{dcsalphabeta4} to:
\begin{eqnarray} \nonumber
 {\rm d}\sigma(\theta|\beta)  &=& 2 \pi   \sum_{k_{_{\rm A}},k_{_{\rm B}}} (2 k_{_{\rm A}}
 +1) (2 k_{_{\rm B}} +1) U^{(k_{_{\rm A}},k_{_{\rm B}})}_{0,0} (\theta) \\  & \times &
 a^{(k_{_{\rm A}})}_{0} a^{(k_{_{\rm B}})}_{0}.
\end{eqnarray}
The coupled-channel quantum calculations to evaluate the scattering matrices are
carried out in full-dimensionality using a modified version of the TwoBC
code~\cite{krems:twobc} and the recently reported full-dimensional
PES for the H$_2$-H$_2$ system~\cite{ZuoH4PES}. This PES was developed by fitting energy points from multi-reference configuration interaction calculations using a permutationally invariant neural network method~\cite{Guo:IRPC16} with the proper electrostatic and long-range dispersion terms.  Details of the scattering calculations are
given in our prior works~\cite{2018_PRL_Croft,2019_JCP_Croft,Bala:JCP09}. For pure
rotational quenching of D$_2(v=2,j=2)$, results are insensitive to the inclusion of additional
rotational or vibrational levels beyond $v$=2 and $j$=4 in the basis set.

\section{results}
\begin{figure}
  \centering
  \includegraphics[width=1.0\linewidth]{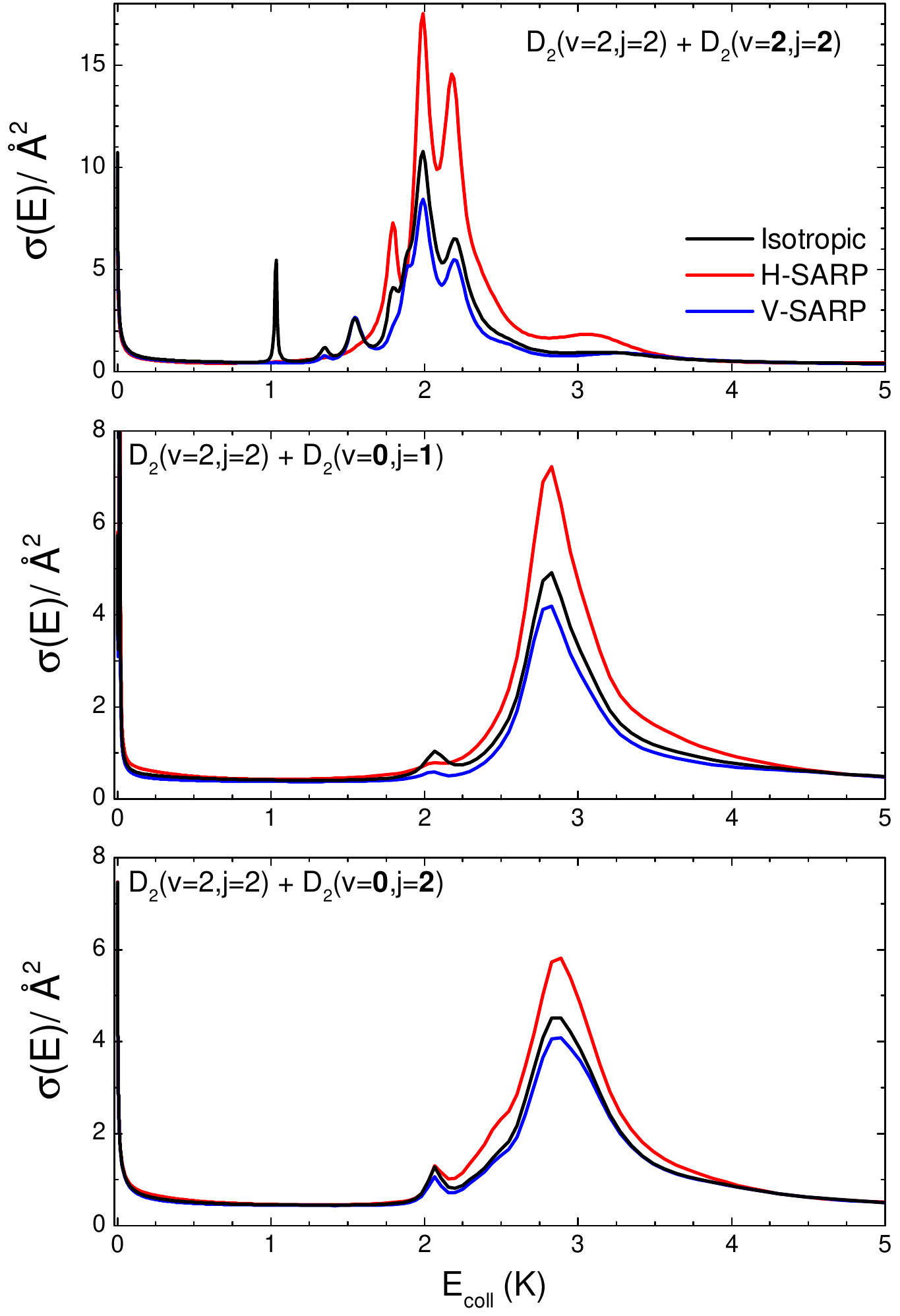}
  \caption{Excitation functions for D$_2$($v'$=2,$j'$=0) production from ($v$=2,$j$=2) +
  ($v$=2,$j$=2) collisions (top panel), ($v$=2,$j$=2) + ($v$=0,$j$=1)(middle panel), and
  ($v$=2,$j$=2) + ($v$=0,$j$=2) (bottom panel). Results  for isotropic preparation is shown in
  black, while those for H-SARP ($\beta$=0$^{\circ}$), and V-SARP ($\beta$=90$^{\circ}$) are
  shown in red and blue, respectively.}\label{Fig1}
\end{figure}
In their experiments, Zhou \textit{et al.}~\cite{Zhou:NC22} used a collimated D$_2$ beam
with a rotational temperature of  $\approx$130\,K (see SI). Using SARP, nearly all
$|v=0,j=0\rangle \equiv  |0\,\,0\rangle$, molecules are transferred to a
$|2\,\,2\rangle$ state. As
a result of the pumping process, the D$_2$ internuclear axes in $|2\,\,2\rangle$ state are
aligned in a chosen direction with respect to the molecular beam axis. Here, we will consider
three possible scenarios: isotropic (no alignment) internuclear axis distribution,
internuclear axis
aligned parallel to the molecular beam axis ($\beta=$0$^{\circ}$ or H-SARP), and internuclear
axis of  D$_2$ $|2\,\,2\rangle$ aligned perpendicular to the molecular beam axis
($\beta=$90$^{\circ}$ or V-SARP). After state preparation, D$_2$ molecules in
$|2\,\,2\rangle$ experience collisions with other D$_2$ molecules in the beam giving rise to a pure
rotational de-excitation to the $|2\,\,0\rangle$ state whose angular distribution is selectively
detected.

Since all the D$_2$ molecules  travel along the molecular beam spanning a
relatively  narrow
velocity distribution, the {\em relative} velocity distribution corresponds to $E_{\rm coll} <$
5~K.  D$_2$ in a $|2\,\,0\rangle$ state can be  produced from inelastic collisions between
either two polarized $|2\,\,2\rangle$ molecules or  between one polarized $|2\,\,2\rangle$
and one unpolarized $|0\,\,1\rangle$ or $|0\,\,2\rangle$ partner. The excitation function
(cross section as a function of $E_{\rm coll}$),  $\sigma(E)$,  for each of these processes are
shown in Figure \ref{Fig1}. For collisions between $|2\,\,2\rangle$ and $|0\,\,1\rangle$ or
$|0\,\,2\rangle$,  $\sigma(E)$  is characterized by a broad resonance peak at $E_{\rm coll}
\sim$ 2.8~K and a smaller peak around 2~K, both  associated with $\ell$=4 (see
Figure ~\ref{FigSres}). Around the resonance,  $\sigma(E)$ is larger for a H-SARP preparation and
slightly smaller for a V-SARP preparation compared to the isotropic case. Away from the
resonance,  $\sigma(E)$ is similar for the three preparations of the $|2\,\,2\rangle$ state.  In
contrast,  $\sigma(E)$  for collisions between two $|2\,\,2\rangle$ molecules displays a
complex resonance structure centered around 2~K,  which are also enhanced by H-SARP
preparation. There is also a sharp resonance at $E_{\rm coll} \sim$ 1~K, that disappears for
both H-SARP and V-SARP polarizations. All these resonances are associated mainly to $\ell$=4
(see Figure ~\ref{FigSres}) and different values of the total angular momentum $J$. Collisions
between two $|2\,\,2\rangle$ molecules that lead to two $|2\,\,0\rangle$ products have a
significantly smaller cross sections, and hence are not considered here.  Irrespective of  the
$\sigma(E)$ shape, although the absolute values for collisions between two $|2\,\,2\rangle$
molecules are  larger, all the three types of encounters have to be considered to account for
the simulation of the  experimental angular distributions.
\begin{figure}
  \centering
  \includegraphics[width=1.0\linewidth]{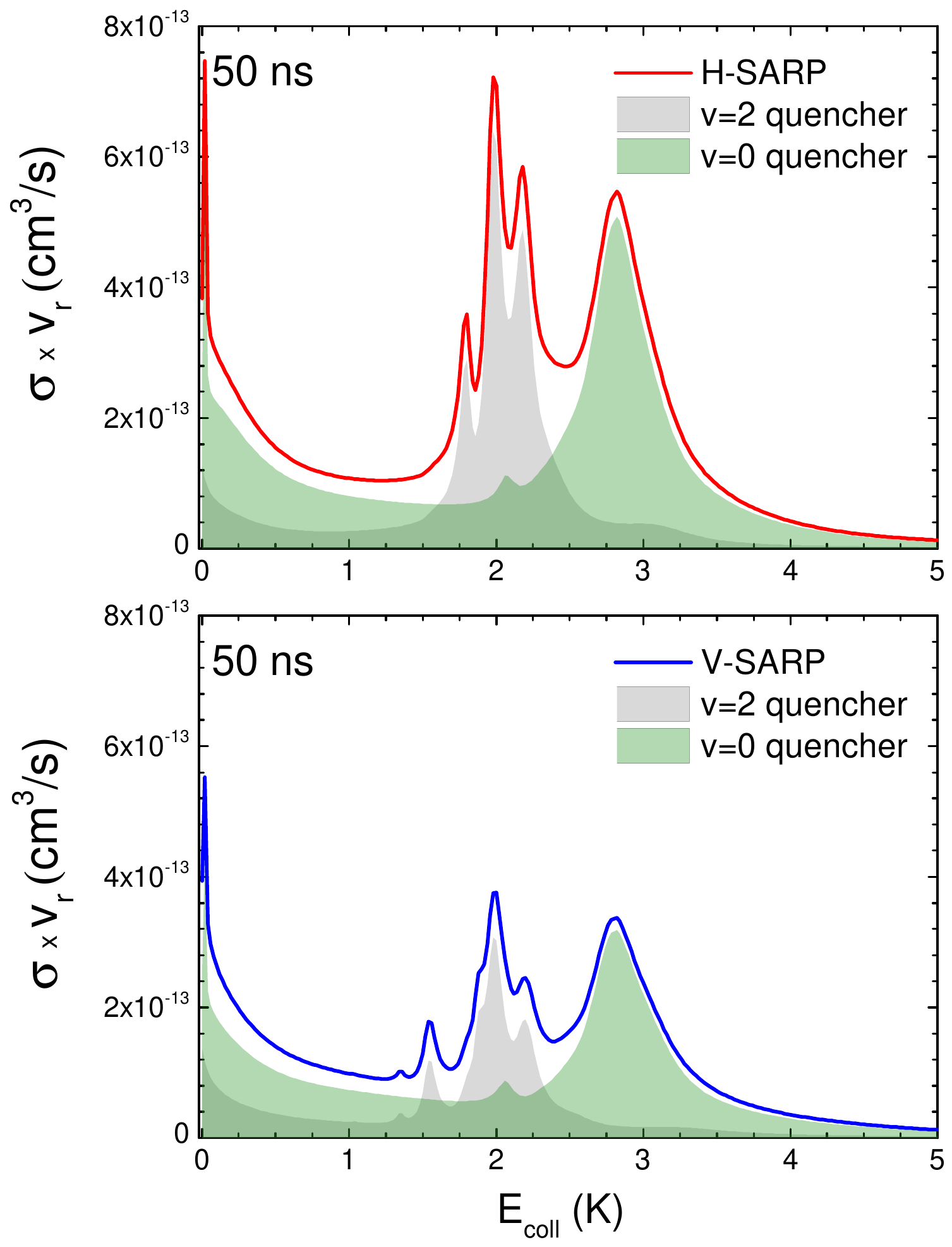}
\caption{ Energy dependent integral rate coefficients multiplied by the experimental
collision energy distribution for a 50 ns SARP-REMPI delay time for the two experimental
preparations:  H-SARP ($\beta$=0$^{\circ}$) (top panel) and V-SARP ($\beta$=90$^{\circ}$)
(bottom panel). The contribution of collisions between two D$_2$($v$=2) molecules is
highlighted in shaded in grey while that from collisions between one D$_2$($v$=2) and one
D$_2$($v$=0) molecule is shown in shaded dark green.}\label{Fig2}
\end{figure}

Figure ~\ref{Fig2} depicts the energy dependent rate coefficients multiplied by the experimental
$E_{\rm coll}$ distribution, such that its integral over $E_{\rm coll}$ is the rate coefficient.
The higher flux for the H-SARP preparation is consistent with its larger cross section compared
to the V-SARP preparation. The different contributions from the $v$=2 and $v$=0 quenchers
are also highlighted.  At $E_{\rm coll}$ within 1.5--2.5\,K, the flux mostly originates from
the resonance features due to ($v$=2) + ($v$=2) collisions, whereas at higher energies the broad
resonance due to ($v$=2) + ($v$=0) collisions prevails.
Overall, the energy
distributions reflect the interplay between resonance features associated with ($v$=2) + ($v$=2) and
($v$=2) + ($v$=0) collision partners, all of them associated to $\ell$=4  (instead of $\ell$=2 as
discussed in Ref.~\cite{Zhou:NC22}), and also show contributions from lower energies,
associated to $\ell$=0 and 1 (see Figure \ref{FigSflux}).
\begin{figure}
  \centering
  \includegraphics[width=1.0\linewidth]{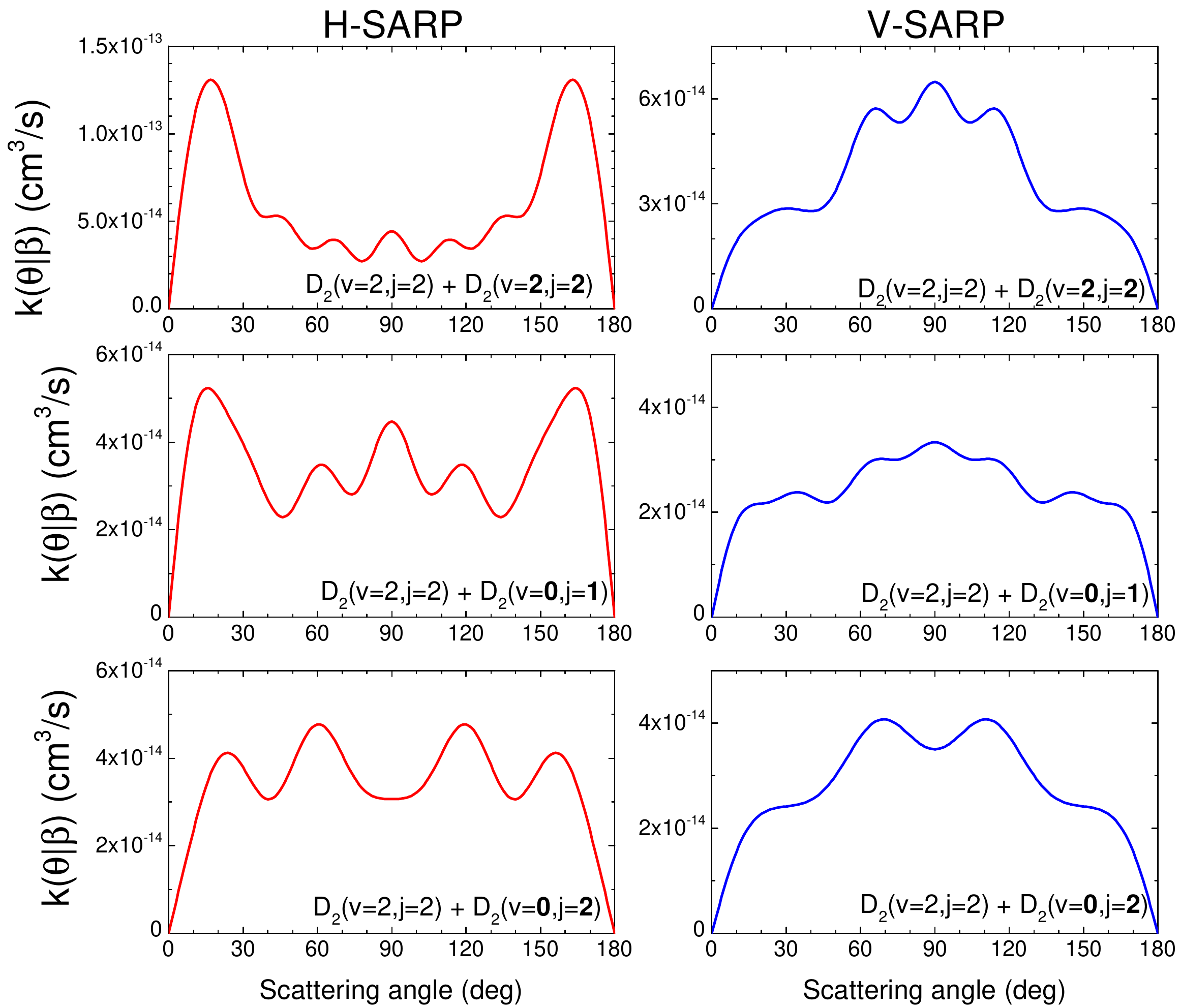}
  \caption{ Velocity-averaged differential rate coefficients for D$_2$($v'$=2,$j'$=0)
  production from ($v$=2,$j$=2) + ($v$=2,$j$=2) collisions (top panel), ($v$=2,$j$=2) +
  ($v$=0,$j$=1)(middle panel), and ($v$=2,$j$=2) + ($v$=0,$j$=2) (bottom panel). Results for a
  H-SARP (V-SARP) preparation are shown in the left (right) panel. Differential rate coefficients
  were symmetrized as discussed in the text.  }\label{Fig3}
\end{figure}

Figure ~\ref{Fig3} shows the computed angular distributions (differential rate coefficients)
convoluted over the experimental velocity distributions for the three collision pairs
considered here and the H-SARP and V-SARP preparations. Since  in the experiments it is not
possible to distinguish between products scattered at $\theta$ or $\pi$-$\theta$ (where
$\theta$ is the scattering angle, that between $\bm{k}$ and $\bm{k'}$), the angular
distributions are symmetrized as in the experiments~\cite{Zhou:NC22}. For H-SARP
preparations between two polarized $|2\,\, 2\rangle$ molecules we observe  prominent
peaks at 15$^{\circ}$ and  165$^{\circ}$.  These peaks are also present for $|2\,\,2\rangle$  +
$|0\,\,1\rangle$ collisions, although in that case, they are not that dominant, and peaks at 60$^{\circ}$, 90$^{\circ}$, and
120$^{\circ}$ also exist. For  $|2\,\,2\rangle$  +  $|0\,\,2\rangle$ collisions the shape is
similar but the magnitude is smaller for the most forward and backward peaks.  The sharp peaks observed for $|2\,\, 2\rangle$ + $|2\,\, 2\rangle$ are a consequence of the simultaneous polarization of both D$_2$ molecules. If, incorrectly, the simulation is carried out just considering polarization of one of the two partners, the shape of the angular distribution is similar to that obtained for $|2\,\, 2\rangle$ + $|0\,\, 1\rangle$ (see Figure \ref{FigSpol}).
For a V-SARP preparation, we obtain a salient  90$^{\circ}$ peak for
$|2\,\,2\rangle$  +  $|2\,\,2\rangle$  collisions that is somewhat  suppressed for
$|2\,\,2\rangle$  +  $|0\,\,1\rangle$ encounters. The angular distribution for
$|2\,\,2\rangle$  + $|0\,\, 2\rangle$  collisions shows a small dip at 90$^{\circ}$ with small
shoulders at each side at 70$^{\circ}$ and 110$^{\circ}$.
\begin{figure}
  \centering
  \includegraphics[width=1.0\linewidth]{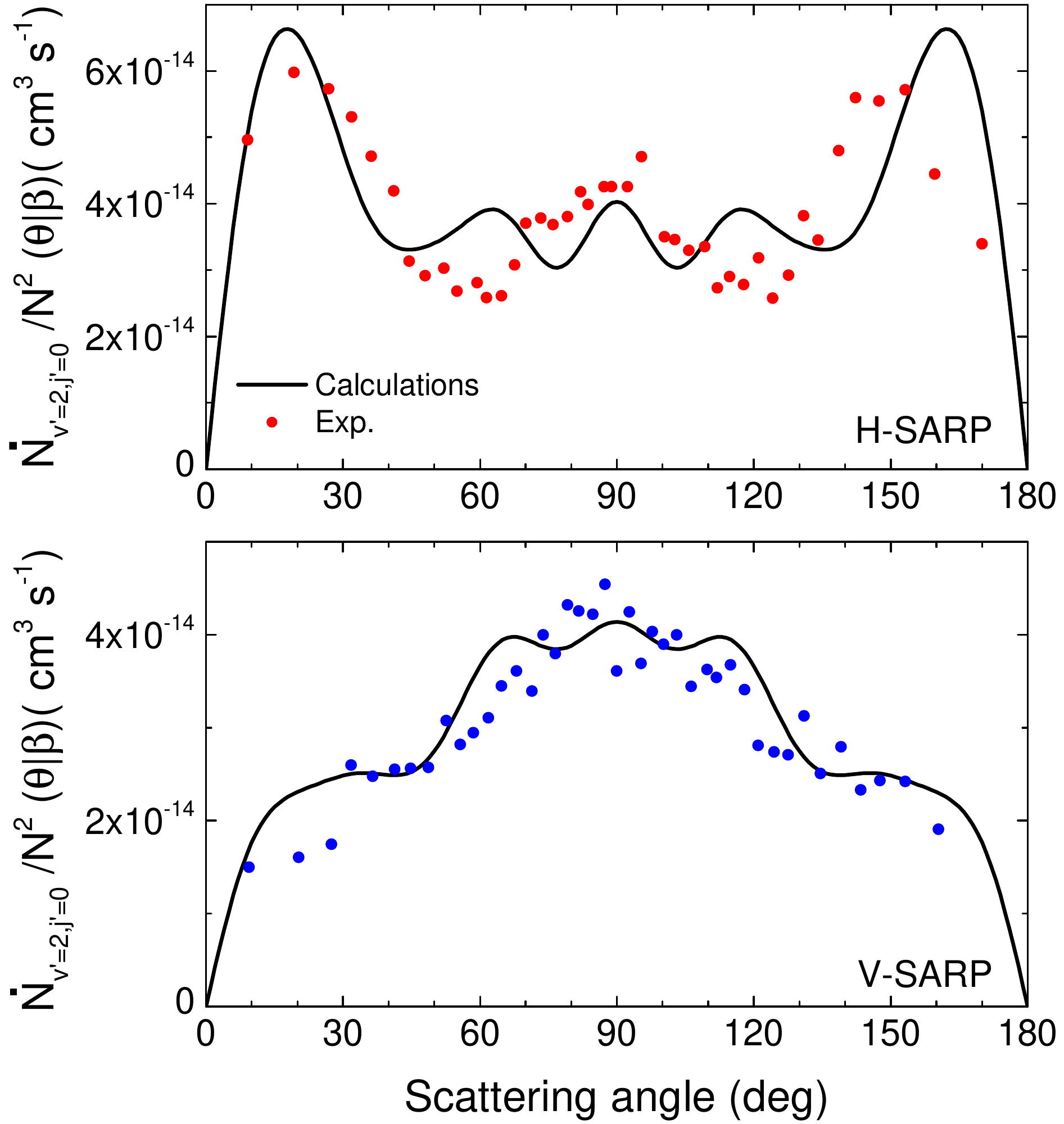}
  \caption{ Initial molecular state and velocity-averaged differential rate coefficients for
  D$_2$($v'$=2,$j'$=0) production normalized by the square of the total density of D$_2$
  (see SI for further details). Results of our calculations are shown in solid curves while
  experimental results of Zhou \textit{et al.}\cite{Zhou:NC22} are shown in dots.  Calculations
  using a H-SARP (V-SARP) preparation are shown in the top (bottom) panels. Differential rate
  coefficients were symmetrized as in the experimental work.  }\label{Fig4}
\end{figure}

Taking into account the populations of the different rovibrational states in the beam, it is
possible to combine the angular distributions depicted in Figure ~\ref{Fig3} and compare with
the experimental angular distributions. Such a comparison is presented in Figure ~\ref{Fig4}.
Note that experiments do not provide absolute values of DCS, so comparison is made on a
relative scale. The agreement between experiment and calculations is good for both H-SARP
and V-SARP. For  H-SARP   our calculations predict that  forward and backward experimental
peaks are caused by $|2\,\,2\rangle  + |2\,\,2\rangle$ collisions while collisions between
$|2\,\,2\rangle + |0\,1\rangle$ and $|0\,2\rangle$  contribute to sideways scattering and, in
particular, to the
smaller sideways peaks. Regarding V-SARP, the experimental signatures  primarily arise from
the $|2\,\,2\rangle$  + $|2\,\,2\rangle$ collisions modulated by small contributions from the
other two collision pairs.

Altogether, our results provide a complete \textit{ab initio} simulation of the experiments of Zhou \textit{et
al.}~\cite{Zhou:NC22} on stereodynamics of bimolecular collisions between two aligned D$_2$ molecules. This is enabled by developing the theory for stereodynamics of aligned-aligned
bimolecular collisions and by considering different collision processes that occur in the
molecular beam. Results presented here based on full-dimensional
coupled-channel scattering
calculations reveal that the angular distribution observed in the experiments of Zhou \textit{et al.}
~\cite{Zhou:NC22} is due to resonance features that arise from different collision partners in
the beam with distinct angular distributions. The formalism presented here is general, and will provide the foundation for describing four-vector correlation in reactive or inelastic aligned molecular collisions in future experiments involving SARP or related techniques.

\section{acknowledgement}

This work was supported in part by NSF grant No. PHY-2110227  (N.B.) and ARO MURI grant
No. W911NF-19-1-0283 (N.B., H.G.). P.G.J.  gratefully acknowledges grant
PID2020-113147GA-I00 funded by MCIN/AEI/10.13039/, and F.J.A. acknowledges   funding
by the Spanish Ministry of Science and Innovation (Grants No. PGC2018-096444-B-I00 and
PID2021-122839NB-I00). J.F.E.C gratefully acknowledges support from the Dodd-Walls Centre for Photonic and Quantum Technologies.

 \bibliographystyle{apsrev4-2}

 \onecolumngrid

\clearpage
\section{Supplementary Information}

\section{Calculation of Initial molecular state and velocity-averaged differential rate
coefficients}

\renewcommand{\figurename}{Fig.}
\renewcommand{\thefigure}{S\arabic{figure}}
\renewcommand{\theequation}{S.\arabic{equation}}
\setcounter{equation}{0}
\setcounter{figure}{0}

 The experiments by  Zhou \textit{et al.}~\cite{Zhou:NC22} made use of a collimated D$_2$
 beam  in which 38\% of the molecules are in  ($v$=0,$j$=0), 36\% in
 ($v$=0,$j$=1), and  24\% in a ($v$=0,$j$=2) state. After the SARP preparation, nearly all
 molecules in  ($v$=0,$j$=0) are pumped  to ($v$=2, $j$=2). Experimentally, it
 was possible to select the
 distribution of the D$_2$($v$=2,$j$=2) internuclear axis in the scattering frame by changing
 the direction of polarization of the laser pulses, defined by the angles $\beta$ and
 $\alpha$ with respect to the scattering frame. Molecules in  D$_2$($v'$=2,$j'$=0) resulting
 from rotationally inelastic collisions between two D$_2$ molecules were probed by (2+1)
 REMPI, and from the velocity distribution of the scattered products, the angular distribution
 could be extracted.  Although the speed of the D$_2$ molecules is about 2 km/s, relative
 collision energies, $E_{\rm coll}$,  are below 5 K, so  D$_2$($v'$=2,$j'$=0) can only be
 produced by quenching of  D$_2$($v$=2,$j$=2). Hence, $\dot{N}_{v'=2,j'=0}(\theta|\beta)$,
 the flux of the scattered D$_2$($v'$=2,$j'$=0) can be calculated as:

\begin{align}\label{eqsim}
  \dot{N} (v'=2,j'=0, \theta|\beta)  \equiv \displaystyle  \frac{{\rm d} N (v'=2,j'=0,
  \theta|\beta)}{{\rm d} t} =   \frac{1}{2}  k_{_{22,22}}(\theta|\beta) \,N_{_{22}} N_{_{22}} +
  k_{_{22,01}}(\theta|\beta) \,N_{_{22}} N_{_{01}} +  k_{_{22,02}}(\theta|\beta)\, N_{_{22}}
 N_{_{02}}
\end{align}
where  $k_{v_{_{\rm A}} j_{_{\rm A}},v_{_{\rm B}} j_{_{\rm B}}}(\theta|\beta)$ are the
differential rate coefficients for collisions between D$_2$ molecules in $(v_{_{\rm A}},
j_{_{\rm A}})$ and $(v_{_{\rm B}}, j_{_{\rm B}})$ states calculated as
\begin{equation}\label{ktheta}
  k(\theta|\beta) =  \bigintss {\rm d}\sigma(\theta|\beta) \sin\theta \, \sqrt{\frac{2 E_{\rm
  coll} }{\mu}} \, f(E_{\rm coll}) \, dE_{\rm coll}
\end{equation}
in which   $\mu$ is the reduced mass of D$_2$+D$_2$ and   $f(E_{\rm coll})$  the
experimental energy distribution. In Eq.~\eqref{eqsim},  $N_{v,j}$ is the molecular density for
the ($v$, $j$) state. The 1/2 factor in $k_{_{22,22}}$ is not to count twice collisions of two
D$_2$ molecules in the same $|2\,\, 2\rangle$ state.

Since the total density of D$_2$ in the experiment is unknown, the results shown in
Figure ~\ref{Fig4} are calculated  by  replacing absolute densities, $N_{v,j}$ by their relative
values $n_{v,j}=  N_{v,j}/ N$, where $N$ is the total D$_2$ density in the molecular
beam.
Hence  the products of the relative densities are $n_{_{22}} n_{_{22}}/2 = 0.2405$, $
n_{_{22}} n_{_{01}} = 0.4556$, and $n_{_{22}} n_{_{02}} = 0.3037$.

In Eq.~\eqref{eqsim} we only consider  collisions in which only one D$_2$($v$=2,$j$=2) is
quenched to  D$_2$($v'$=2,$j'$=0) while its collision partner is unperturbed. Double
relaxation collisions, albeit possible, are associated with much lower cross sections and  their
contribution to the experiment is negligible.

Collisions between ($v$=2,$j$=2) and ($v$=0,$j$=1) involve one $o-$D$_2$ and one
$p-$D$_2$ molecules, which are distinguishable. However, collisions between ($v$=2,$j$=2)
and ($v$=0, 2, $j$=2) involve two indistinguishable $o-$D$_2$ molecules. To carry out
calculations for two indistinguishable molecules, the wave function was symmetrized with
respect to the exchange-permutation symmetry of the molecules,\cite{Bala:JCP09}  and the
statistically weighted sum of the exchange-permutation symmetrized cross sections is given
by \cite{HuoGreen:JCP96}
\begin{equation}\label{sigmanuclearspin}
  \sigma_{j'_{\rm A} j'_{\rm B}, j_{_{\rm A}} j_{_{\rm B}}} = w^+  \sigma^+_{j'_{\rm A} j'_{\rm
  B}, j_{_{\rm A}} j_{_{\rm B}}} + w^-  \sigma^-_{j'_{\rm A} j'_{\rm B}, j_{_{\rm A}} j_{_{\rm
  B}}}
\end{equation}
where $w^{\pm}$ are the statistical weights of nuclear spin states associated with even or
odd exchange symmetries of the two identical D$_2$ nuclei.
In the particular case of collisions between two $o-$D$_2$ molecules \cite{Johnson:JCP79},
\begin{equation}
  w^+ = \frac{21}{36}, \quad w^- = \frac{15}{36}.
\end{equation}

An equation similar to \eqref{sigmanuclearspin} holds for the calculation of any other
observable when indistinguishable particles are involved, such as ${\rm
d}\sigma(\theta|\beta, \alpha)$:
\begin{eqnarray}\label{sigmanuclearspin_DCS}
  {\rm d}\sigma(\theta|\beta,\alpha) = w^+ {\rm d}\sigma^+(\theta|\beta,\alpha) +  w^-  {\rm
  d}\sigma^-(\theta|\beta,\alpha)
\end{eqnarray}
where the scattering amplitudes are given by
\begin{eqnarray} \label{scatamplpar}
  F^{\pm}_{m'_{\rm A} m'_{\rm B} \,m_{_{\rm A}} m_{_{\rm B}}}(\theta) &=& \frac{1}{2 i k}
  \sum_{J} (2J +1) d^J_{m'_{\rm A} + m'_{\rm B}, m_{_{\rm A}} + m_{_{\rm B}} }(\theta) \,
  S^{J,\pm}_{m'_{\rm A} m'_{\rm B} \,m_{_{\rm A}} m_{_{\rm B}}}(E).
\end{eqnarray}

In the particular case of ($v$=2,$j$=2) + ($v$=0,$j$=2) collisions, $S^{J,+}_{m'_{\rm A}
m'_{\rm B} \,m_{_{\rm A}} m_{_{\rm B}}}(E) \sim S^{J,-}_{m'_{\rm A} m'_{\rm B} \,m_{_{\rm
A}} m_{_{\rm B}}}(E)$. As discussed by Huo and Green \cite{HuoGreen:JCP96} this implies
that the two collision partners are virtually distinguishable (there is no interference between
the direct and exchange mechanisms, the latter being negligible). On the contrary, for the
($v$=2,$j$=2) + ($v$=2,$j$=2) collisions $S^{J,+}$ and $S^{J,-}$ are very different, which is
indicative of a strong interference between direct and exchange mechanisms.

\clearpage

\section{Contribution of $\ell$ partial waves to the excitation functions}

In this section we present the decomposition of the integral cross sections into the various
contributions from  the orbital angular momentum  $\ell$-partial waves.

\begin{figure}[h!]
  \centering
  \includegraphics[width=0.5\linewidth]{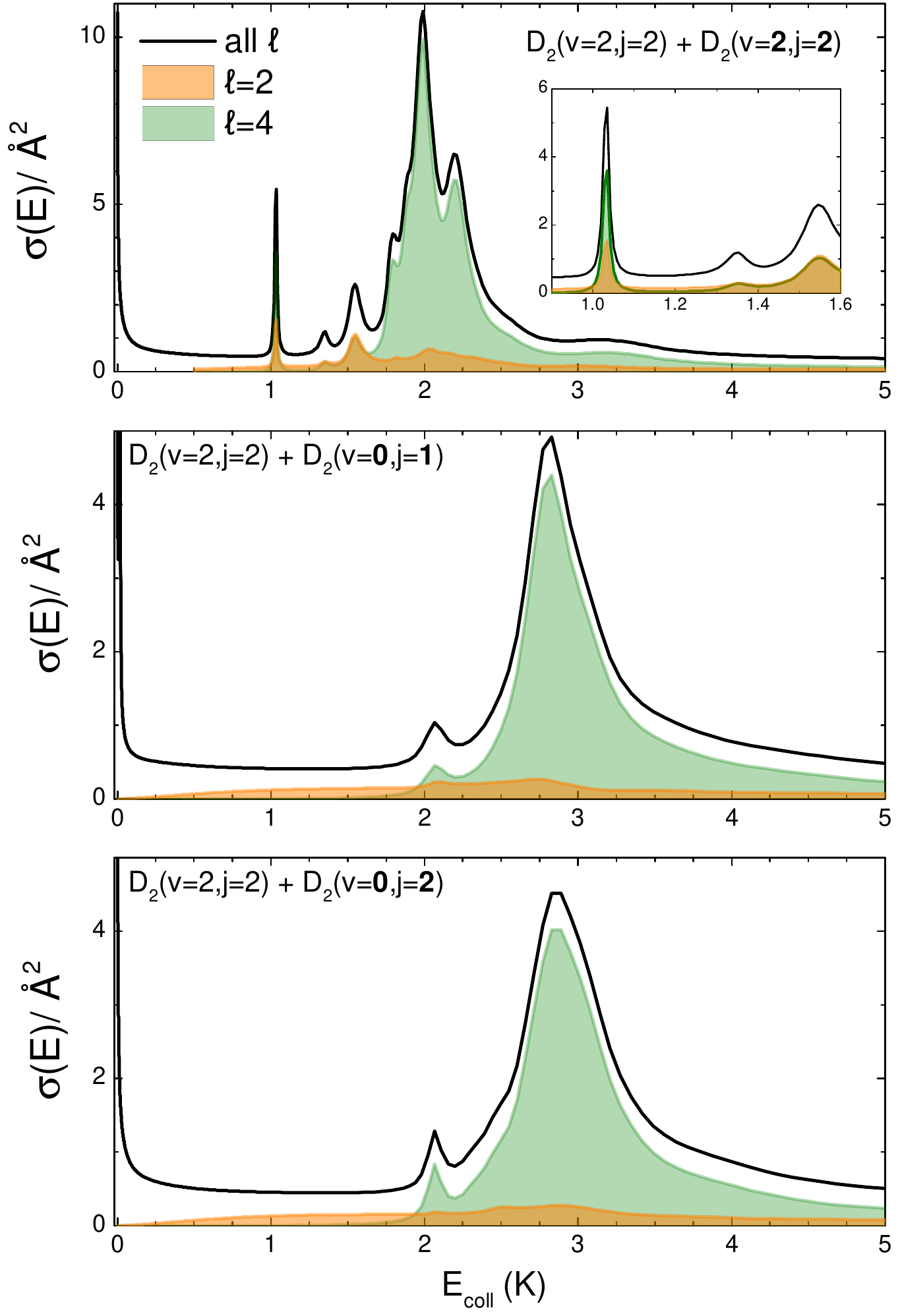}
  \caption{ Isotropic excitation functions for D$_2$($v'$=2,$j'$=0) production from
  ($v$=2,$j$=2) + ($v$=2,$j$=2) collisions (top panel), ($v$=2,$j$=2) + ($v$=0,$j$=1)(middle
  panel), and ($v$=2,$j$=2) + ($v$=0,$j$=2) (bottom panel). Results for all $\ell$
  are   shown by black curves, while those including only $\ell$=2 and $\ell$=4 are shaded in
  orange  and green, respectively. As it is apparent from the figure, resonances are caused mainly by
  $\ell$=4, although there is a small contribution from  $\ell$=2.   }\label{FigSres}
\end{figure}

\begin{figure} [h!]
  \centering
  \includegraphics[width=0.7\linewidth]{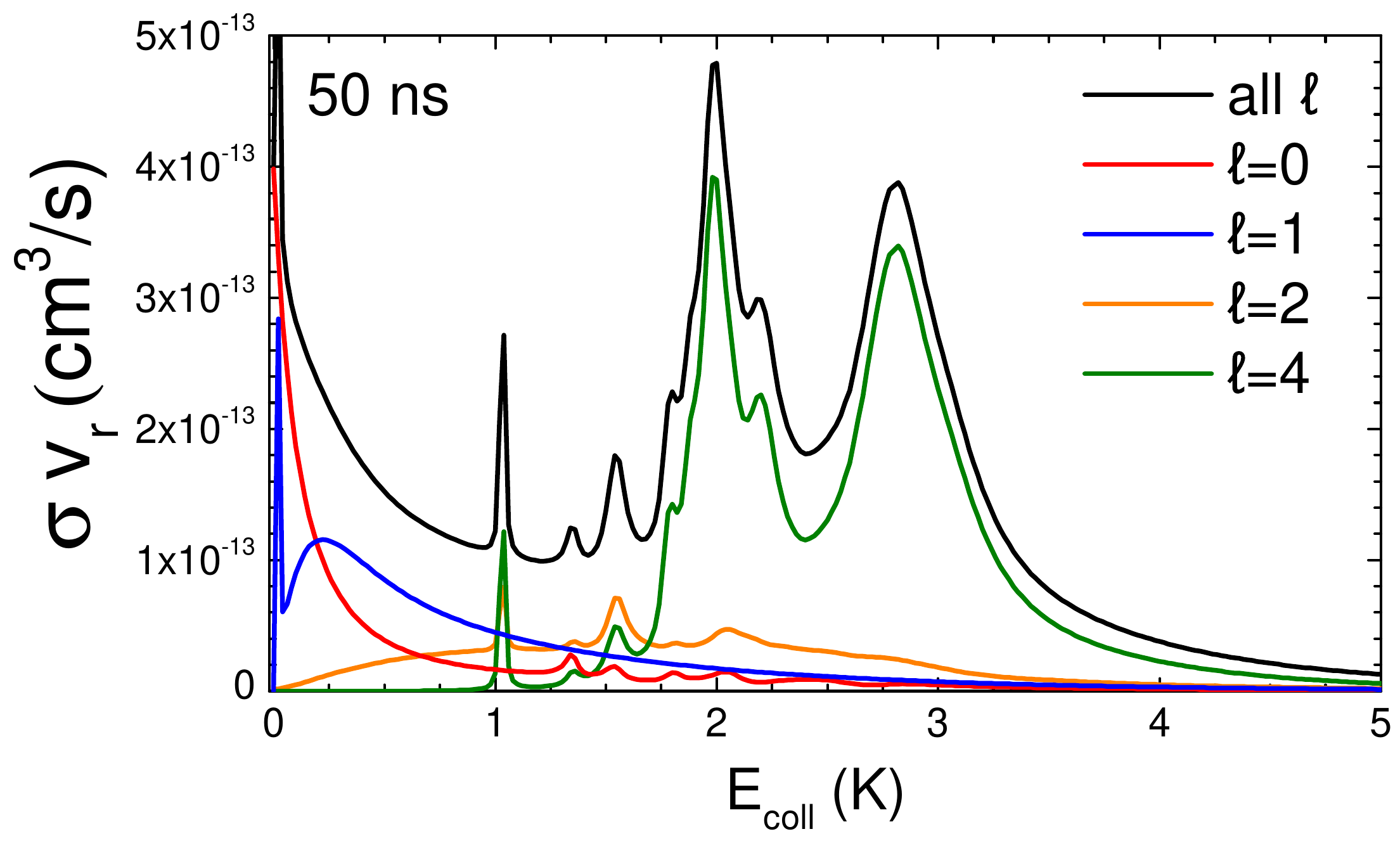}
  \caption{Energy dependent flux weighted with energy distribution for an isotropic
  preparation comprising all the possible quenchers. The contributions from  different
  $\ell$ partial waves are also shown.  }\label{FigSflux}
\end{figure}

Figure \ref{FigSres} displays the contributions from $\ell=2$ and $\ell=4$ to the isotropic
(unaligned D$_2$ molecules)  integral cross sections (excitation functions)  for D$_2(v=2,j=2)$
+ D$_2(v=2,j=2) \to$ D$_2(v=2,j=0)$ + D$_2(v=2,j=2)$ (top panel), D$_2(v=2,j=2)$ +
D$_2(v=0,j=1) \to$ D$_2(v=2,j=0)$ + D$_2(v=0,j=1)$ (middle panel), and D$_2(v=2,j=2)$ +
D$_2(v=0,j=2) \to$ D$_2(v=2,j=0)$ + D$_2(v=0,j=2)$ (lower panel). As can be seen, the
prevailing contribution in all cases is that from $\ell=4$. Specifically, the resonances for
quenching with D$_2(v=0, j=1,2)$ are exclusively due to the $\ell =4$.
\newpage
Figure~\ref{FigSflux} shows  the energy dependent flux (multiplied by the relative collision
energy distribution at 50 ns)  in the isotropic case including the contributions from all the
possible quenchers. The total flux is broken down into the various $\ell$ contributions.
As can be  seen, $\ell$=4 contributes mostly to the scattering, with $\ell$=0,
and 1 becoming prominent at $E_{\rm coll} <$1 K.

\section{Differences between three- and four-vector correlations}

\begin{figure}
  \centering
  \includegraphics[width=1.0\linewidth]{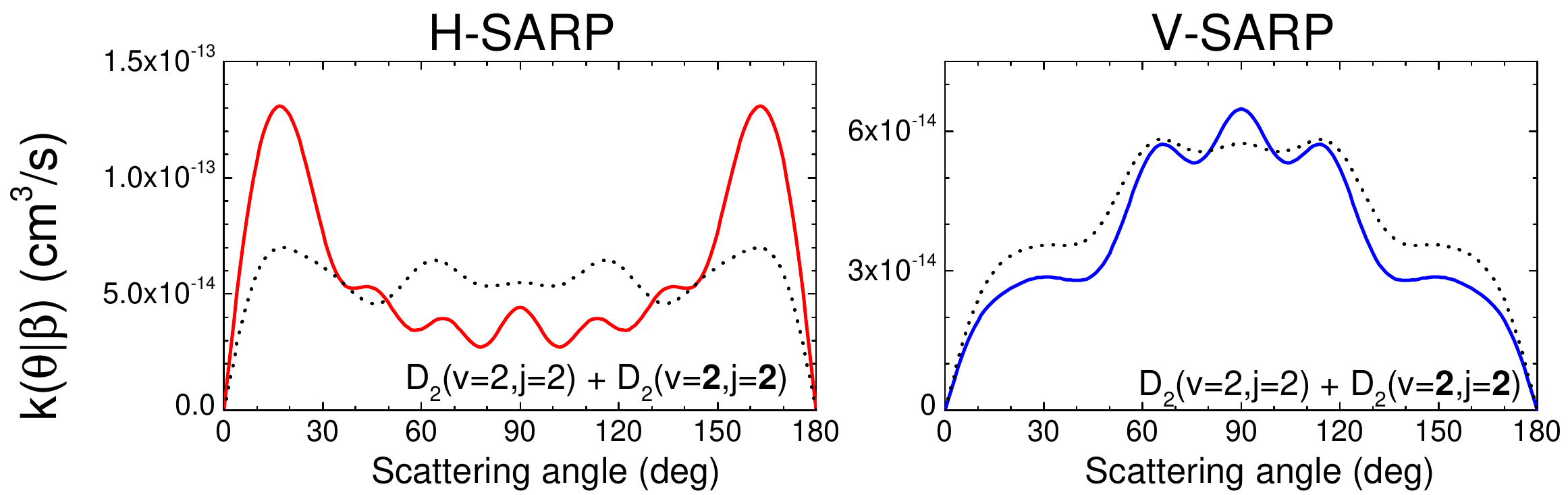}
  \caption{ Velocity-averaged differential rate coefficients for D$_2$($v'$=2,$j'$=0)
  production from ($v$=2,$j$=2) + ($v$=2,$j$=2) collisions considering the full (4-vector) correlations (solid lines) and  a scenario where only two of the two partners is polarized (dashed line).  }\label{FigSpol}
\end{figure}

To properly account for the experiments, polarization of both  molecules should be considered in the collisions between two indistinguishable D$_2$ ($v$=2,$j$=2)  molecules, and 4-vector \{${\bm k}$--${\bm j}_{_{\rm A}}$\!\!--$ {\bm
j}_{_{\rm B}}$\!\!--${\bm k}'$\} correlations are needed. To assess the importance of the 4-vector correlations,  Figure  \ref{FigSpol} shows the effect of simulating the velocity-averaged differential rate coefficients using only three-vector PDDCSs (i.e., a scenario in which only one of the two partners is polarized). As can be seen, for V-SARP there is almost no difference, while for H-SARP we observe how the two prominent peaks at 15$^{\circ}$ and 165$^{\circ}$ are significantly less intense when 4-vector PDDCSs are not included. Actually, the neglect of 4-vector correlations leads to velocity-averaged differential rate coefficients with a similar shape as those obtained for  D$_2$ ($v$=2, $j$=2) +  D$_2$ ($v$=2, $j$=1,2) collisions.

\end{document}